\documentclass{desyproc}

\usepackage{bm}
\usepackage{graphicx}
\usepackage{dcolumn}
\usepackage{epsfig}
\usepackage[cp1251]{inputenc}
\usepackage[english]{babel}
\usepackage{euscript}
\usepackage{amstext}
\usepackage{amsmath}
\usepackage{amssymb}
\usepackage{subfigure}
\usepackage{wrapfig}
\DeclareGraphicsExtensions{.pdf,.jpg,.eps}

\begin{document}
\title{Physics Potential of the CMS CASTOR Forward Calorimeter}

%
\author{{\slshape Dmytro Volyanskyy}  for the CMS Collaboration\\[1ex]
DESY, Notketra{\ss}e 85, 22607 Hamburg, Germany }

\contribID{xy}  
\confID{1964}
\desyproc{DESY-PROC-2010-01}
\acronym{PLHC2010}
\doi            

\maketitle

\begin{abstract}

The CASTOR calorimeter is a detector covering the very forward region of the CMS experiment at the LHC. 
It surrounds the beam pipe with $14$ longitudinal modules each of which consisting of $16$ azimuthal sectors 
and allows to reconstruct shower profiles, separate electrons and photons from hadrons 
and search for phenomena with anomalous hadronic energy depositions. 
The physics program that can be performed with this detector includes a large variety 
of different QCD topics. In particular, the calorimeter is supposed to contribute to studies 
of low-$x$ parton dynamics, diffractive scattering, multi-parton interactions and 
cosmic ray related physics in proton-proton and heavy-ion collisions. 
The physics capabilities of this detector are briefly summarized in this paper.

\end{abstract}

\section{Detector Overview}
The CASTOR~(Centauro And STrange Object Reseacrh) detector is located at a distance of $14.4$~m from the CMS interaction point 
right behind the Hadronic Forward~(HF) calorimeter and the T2, a tracking station of the TOTEM experiment, 
covering the pseudorapidity region  $-6.6<\eta<-5.2$. 
This is a quartz-tungsten Cerenkov sampling calorimeter.
%
\begin{wrapfigure}{r}{0.5\textwidth}
  \vspace{-20pt}
  \begin{center}
    \includegraphics[width=0.48\textwidth]{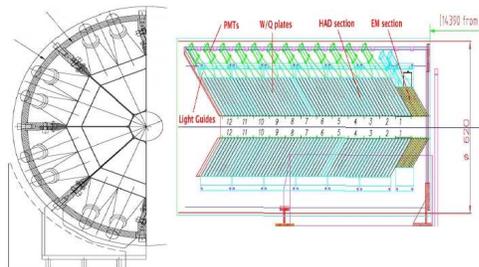}
  \end{center}
  \vspace{-20pt}
  \caption{\sl Sketch of the CASTOR calorimeter: front view~(left) and longitudinal cross section~(right).}
  \vspace{-10pt}
\end{wrapfigure}
That is, it is made of repeating layers~(arranged in a sandwich structure) of quartz and tungsten plates. 
The former is used as the active material because of its radiation hardness, 
while the latter serves as the absorber medium providing the smallest possible shower size. 
The signal in CASTOR is produced when charged shower particles
pass through the quartz plates with the energy above the Cerenkov threshold ($190$~keV for electrons).
The generated Cerenkov light is then collected by air-code light guides,
which are transmitting it further to photo-multipliers tubes PMTs. 
These devices produce signals proportional to the amount of light collected. 
As can be seen in Figure~1, the detector plates are tilted at $45^{0}$ w.r.t. the beam axis
to maximaize the Cerenkov light output in the quartz. The CASTOR detector is a compact calorimeter with the physical size of about $\rm 65~cm \times 36~cm \times 150~cm$
and having no segmentation in $\eta$.  It is embedded into a skeleton, which is made of stainless steel.
The detector consists of $14$ longitudinal modules, each of which comprises $16$ azimuthal sectors
that are mechanically organized in two half calorimeters.
First $2$ longitudinal modules form the electromagnetic section, while the other $12$ modules form the hadronic section.
In the electromagnetic section, the thicknesses of the tungsten and quartz plates are $5.0$ and $2.0$~mm, respectively. 
The corresponding thicknesses in the hadronic section are twice as large as in the electromagnetic section.
With this design, the diameter of the showers of electrons and positrons produced by hadrons
is about one cm, which is an order of magnitude smaller than in other types of calorimeters.
The detector has a total depth of $10.3$ interaction lengths and includes $224$ readout channels.
It should be noted that the final CASTOR design is the result of three test beam campaigns and numerous Monte Carlo simulations.
After the completion of the detector construction in the spring of 2009, the calorimeter
has been successfully installed and commissioned in the summer of 2009. 

\section{The CASTOR physics capabilities}

Because of its pseudorapidity coverage, CASTOR significantly expands 
the CMS capability to investigate physics processes occurring at very low polar angles and so, providing a valuable tool 
to study low-$x$ QCD, diffractive scattering, multi-parton interactions and underlying event structure. 
Another CASTOR objective is to search for exotic objects with unusual longitudinal shower profile, 
several of which have been observed in cosmic ray experiments.

\subsection{Low-x QCD}
A study of QCD processes at a very low parton momentum fraction $x=p_{\rm parton}/p_{\rm hadron}$
is a key to understand the structure of the proton, whose gluon density
is poorly known at very low values of $x$. 
At the LHC the minimum accessible $x$ in proton-proton~($pp$) collisions
decreases by a factor of about $10$ for each $2$ units of rapidity.
This implies that a process with a hard scale of $Q \sim 10$~GeV
and within the CASTOR acceptance can probe quark densities down $x \sim 10^{-6}$~[1], that has never been achieved before.
Such processes include the production of forward jets and Drell-Yan electron pairs. 
The latter occurs via the $qq\rightarrow\gamma^{*} \rightarrow e^{+}e^{-}$ reaction 
within the acceptance of CASTOR and TOTEM-T2 station, whose usage is essential for detecting these events. 
Measurements of Drell-Yan events can also be used to study QCD saturation effects -- 
the effects of rising of the gluon density in the proton with decreasing values of $x$,
that have been firstly observed at HERA. It was found that the Drell-Yan production cross section 
is suppressed roughly by a factor of $2$ when using a PDF with saturation effects compared to one without.
Another way to constrain the parton distribution function~(PDF) of the proton at low~$x$ is provided 
by measuring forward jets in CASTOR that will enable to probe the parton densities down $10^{-6}$. 
Moreover, this allows to gain information on the full QCD evolution to study high order QCD reactions. 
Apart from that, it has been found that a BFKL like simulation, for which the gluon ladder is ordered in $x$, 
predicts more hard jets in the CASTOR acceptance than the DGLAP model that assumes strong ordering 
in the transverse momentum $k_{\rm T}$ and random walk in $x$. 
Therefore, measurements of forward jets in CASTOR can be used as a good tool
to distinguish between DGLAP and non-DGLAP type of QCD evolution.
Furthermore, CASTOR in combination with HF can be used to measure Mueller-Navalet dijet events,
which are characterized by two jets with similar $p_{\rm T}$ but large rapidity separation. 
By measuring Mueller-Navalet dijets in CASTOR one can probe BFKL-like dynamics and small-$x$ evolution.

\subsection{Diffraction}
A good way to study the perturbative QCD and the hadron structure is provided by 
diffractive $pp$ interactions (where one or both the colliding protons stay intact) 
via measurements of the cross sections for diffractive $W$, $Z$, jet or heavy quark productions.
The CASTOR calorimeter is, in particular, a very useful tool to measure 
the single-diffractive productions of $W$ and dijets in $pp$ collisions 
($pp \rightarrow p X$ reaction, where $X$ is either a $W$ boson or a dijet system). 
These are hard diffractive processes that are sensitive to the quark and gluon content of the low-$x$ proton PDFs, correspondingly. 
A selection of such events can be performed using the multiplicity distributions of tracks
in the central tracker and calorimeter towers in HF plus CASTOR exploiting the fact that diffractive events
on average have lower multiplicity in the central region and in the $``$gap side$"$ than non-difractive ones.
Feasibility studies to detect the single-diffractive productions of $W$~[2]  and dijets~[3]
have shown that the diffractive events peak in the regions of no activity in HF and CASTOR.

\subsection{Multi-parton interactions and underlying event structure}
Measurements of energy deposits in the CASTOR acceptance should significantly improve our understanding of the 
multi-parton interactions~(MPI) and underlying event~(UE) structure. The latter is  
an unavoidable background to most collider observables, whose understanding is essential 
for precise measurements at the LHC. It consists of particles 
arising from the beam-beam remnants and from MPI. 
The MPI arise in the region of small-$x$ where parton densities are large so that 
the likelihood of more than one parton interaction per event is high.
According to all QCD models, the larger the collision energy the greater the contribution 
from MPI to the hard scattering process. However, this dependence is currently weakly known.
Measurements of the forward energy flow by means of CASTOR will allow to discriminate 
between different MPI models, which vary quite a lot. 
Furthermore, measurements of forward particle production
in $pp$ and Pb-Pb collisions at LHC energies with CASTOR should help to significantly improve
the existing constraints on ultra-high energy cosmic ray models.

\section{Conclusion}
The CASTOR calorimeter is a valuable CMS subcomponent allowing 
to perform a very rich physics program. The detector is fully integrated in the CMS readout 
and currently take collision data. Its first physics results are currently under preparation.

\section{Acknowledgments} 
I am very thankful to Hannes Jung, Kerstin Borras and many other colleagues working in
the CMS forward physics community for fruitful discussions, suggestions and encouragements.


\begin{footnotesize}

\end{footnotesize}


\end{document}